\documentclass{article} 
\usepackage{subfigure}                           
\usepackage{slashed}
\usepackage{graphicx}
\usepackage{epstopdf} 
\usepackage{verbatim}
\usepackage{multicol}
\usepackage{natbib}
\usepackage{amsmath}
\usepackage{amssymb}
\usepackage[top=2.5cm, bottom=2.5cm, left=2cm, right=2cm]{geometry}
\usepackage{multirow}
\usepackage{sectsty}
\usepackage{color}
\usepackage{hyperref} 

\usepackage[normalem]{ulem}

\sectionfont{\large}

\begin{document}
\title{
A comment on ``Amplification of endpoint structure for new particle mass measurement at the LHC''
}
\author{A. J. Barr${}^{a}$, C. Gwenlan${}^{a}$, C.G. Lester${}^{b}$, C. J. S. Young${}^{a}$. \\ ${}^{a}$\small{Dept. of Particle Physics, Oxford University}\\ ${}^{b}$\small{Department of Physics, Cavendish Laboratory, JJ Thomson Avenue, Cambridge, CB3 0HE, United Kingdom}}
\date{\small\today}
\maketitle

\begin{abstract}
We present a comment on the kinematic variable $m_{CT2}$ recently proposed in \cite{Cho:2009ve}. The variable is designed to be applied to models such as $R$-parity conserving Supersymmetry (SUSY) when there is pair production of new heavy particles each of which decays to a single massless visible and a massive invisible component. It was proposed in \cite{Cho:2009ve} that a measurement of the peak of the $m_{CT2}$ distribution could be used to precisely constrain the masses of the SUSY particles. We show that when Standard Model backgrounds are included in simulations, the sensitivity of the $m_{CT2}$ variable to the SUSY particle masses is more seriously impacted for  $m_{CT2}$ than for other previously proposed variables.
\end{abstract}

\begin{multicols}{2}

If new physics is discovered at the Large Hadron Collider (LHC) \cite{Evans:2008zzb} we will wish to measure the masses of any new particles discovered. Several methods of measuring the masses of such new particles have been suggested (for a recent review see \cite{Barr:2010zj}). 

Here we comment on the mass measurement variable $m_{CT2}$ recently proposed by Cho {\it et al.}~\cite{Cho:2009ve}. The variable introduced in that paper was employed to constrain the participating particles' masses for events characterised by pair production of identical-mass heavy particles, $Y$, each of which decays to a massive invisible $\tilde{\chi}^0_1(\slashed{p})$ and a single massless visible $v(p)$ daughter, where the symbols in brackets label the momenta.

The experimental signature for this process 
\begin{eqnarray}
q\bar{q},gg & \rightarrow & Y^{(1)}\bar{Y}^{(2)} \nonumber\\
Y^{(i)} & \rightarrow & \tilde{\chi}^0_1(\slashed{p}^{(i)}) + v(p^{(i)}) \label{eq:topology}
\end{eqnarray}
is therefore two visible particles (in this paper we assume these to be jets and those jets to be massless) and large missing momentum.

Cho {\it et al.} suggest constraining the masses by measuring the distribution of the variable \cite{Cho:2009ve} 
\begin{eqnarray}
&&\!\!\!\!\!\!\!\!\!\!m_{CT2}(p^{(1)}_T,p^{(2)}_T,\slashed{p}_T,\chi)\label{eq:mct2def}\\
&&\!\!\!\!\!\!\!\!\!\!\,\,\equiv \underset{\sum\slashed{q}^{(i)}_T = \slashed{p}_T}{\textrm{min}} \left(\textrm{max}\left(m_{CT}(p^{(1)}_T,\slashed{q}_T^{(1)}),m_{CT}(p^{(2)}_T,\slashed{q}_T^{(2)})\right)\right) .\nonumber
\end{eqnarray}
This variable is a novel amalgam of two methods previously described in the literature.
The procedure of minimising the larger of two quantities --- over all partitions of the invisible particles' transverse momenta consistent with the missing transverse momentum $\slashed{p}_T$ ---
 has its origins in the `\textbf{\large s}transverse mass' of \cite{Lester:1999tx,Barr:2003rg}. 
However, whereas \cite{Lester:1999tx,Barr:2003rg} take the larger of the two transverse masses $m_T$ for each $Y$ decay, in \eqref{eq:mct2def} the quantities being evaluated at each possible partition of $\slashed{p}_T$ are the {\em con}transverse mass functions $m_{CT}$.
These are defined (for $m_v=0$) by\footnote{$m_{CT}$ 
was originally proposed in \cite{Tovey:2008ui} for the case where both momenta correspond to {\em visible} particles
and was motivated by particular invariance properties under {\em back-to-back} boosts in the transverse plane.
The use of $m_{CT}$ for the case where one input corresponds to a visible particle, but the other represents the hypothesised momentum of an invisible particle was an innovation of \cite{Cho:2009ve}. 
} 
\begin{equation}\label{eq:mctdef}
m_{CT}^2(p_T,\slashed{q}_T,\chi)=\chi^2+2 E(p_T,0)E(\slashed{q}_T,\chi)+2{p}_T \cdot {\slashed{q}}_T,
\end{equation}
where the transverse energy of a particle with transverse momentum $p_T$ and mass $m$ is given by
$E(p_T,m) = \sqrt{p_T^2 + m^2}$,
and $\chi$ is a trial value for the ({\it a priori} unknown) mass of the invisible particle. 
The $+$ sign in front of the inner product of the momenta in \eqref{eq:mctdef} distinguishes $m_{CT}$ from the usual transverse mass.

The already bloated dictionary of transverse mass variable names is stretched nearer to (or perhaps beyond) breaking point by the addition of the term {\em con\textbf{\large s}transverse mass} for the quantity defined in \eqref{eq:mct2def}.

It was suggested in  \cite{Cho:2009ve} that by measuring the kinematic end-point of the $m_{CT2}$ distribution it should be possible to determine the mass of the parent particle $Y$ accurately. This assertion is based on the observation \cite{Cho:2009ve} that under the condition\footnote{Colloquially known as the condition of no `upstream transverse momentum'.}
\begin{equation}\label{eq:noutm}
\slashed{p}_T = - p_T^{(1)} - p_T^{(2)}.
\end{equation}
the distribution of $m_{CT2}$ is highly peaked at its maximal value (under that same condition),
\begin{equation}\label{eq:mct2max}
\left(m_{CT2}^{\textrm{max}}\right)^2=\chi^2+2(E_T(p_0,\chi)|p_0|-|p_0|^2),
\end{equation}
where $|p_0|$ is the absolute momentum of the daughter particle in the rest frame of the parent\footnote{For a two body decay $|p_0|=\left({(m_{Y}^{\textrm{true}})}^2-{(m_{\chi}^{\textrm{true}})}^2\right)/2m_{Y}^{\textrm{true}}$ where $m_Y^{\textrm{true}}$ and $m_{\chi}^{\textrm{true}}$ are the true masses of the parent and daughter SUSY particles respectively.}.

The central observation of \cite{Cho:2009ve} is that the $m_{CT2}$ distribution has a sharply defined Jacobian peak at its kinematic endpoint \eqref{eq:mct2max}, so a good measurement of that endpoint position could provide a precise constraint on the parent particle mass. 
This statement is founded on the reasonable expectation that the systematic uncertainty in fitting a sharp peak ought to be smaller than in fitting other distributions which tend to have smaller numbers of events near their kinematic endpoints.

We perform simulations similar to those in \cite{Cho:2009ve} but including the most important Standard Model backgrounds. We find that those backgrounds are peaked in the same region as the signal, so play a much more significant role for the process of interest \eqref{eq:topology} than might be inferred from \cite{Cho:2009ve}. 
The end-point value \eqref{eq:mct2max} is relatively insensitive to the physical parameter $|p_0|$, so a rather precise measurement of the peak position would be required to constrain the parent particle masses. We suggest that fitting the endpoint position with the required precision is likely to be difficult when the systematic effect from uncertain residual Standard Model backgrounds is taken into account.

\begin{figure*}[t]
\begin{center}
\begin{minipage}{0.9\linewidth}
  \setcounter{subfigure}{0} 
  \subfigure[$m_{CT2}(\chi=300\,\textrm{GeV})$ distribution with just the dijet cuts applied.]{\label{fig:MCT2dijetcuts}\includegraphics[width=77mm,trim=0 21 0 21,clip]{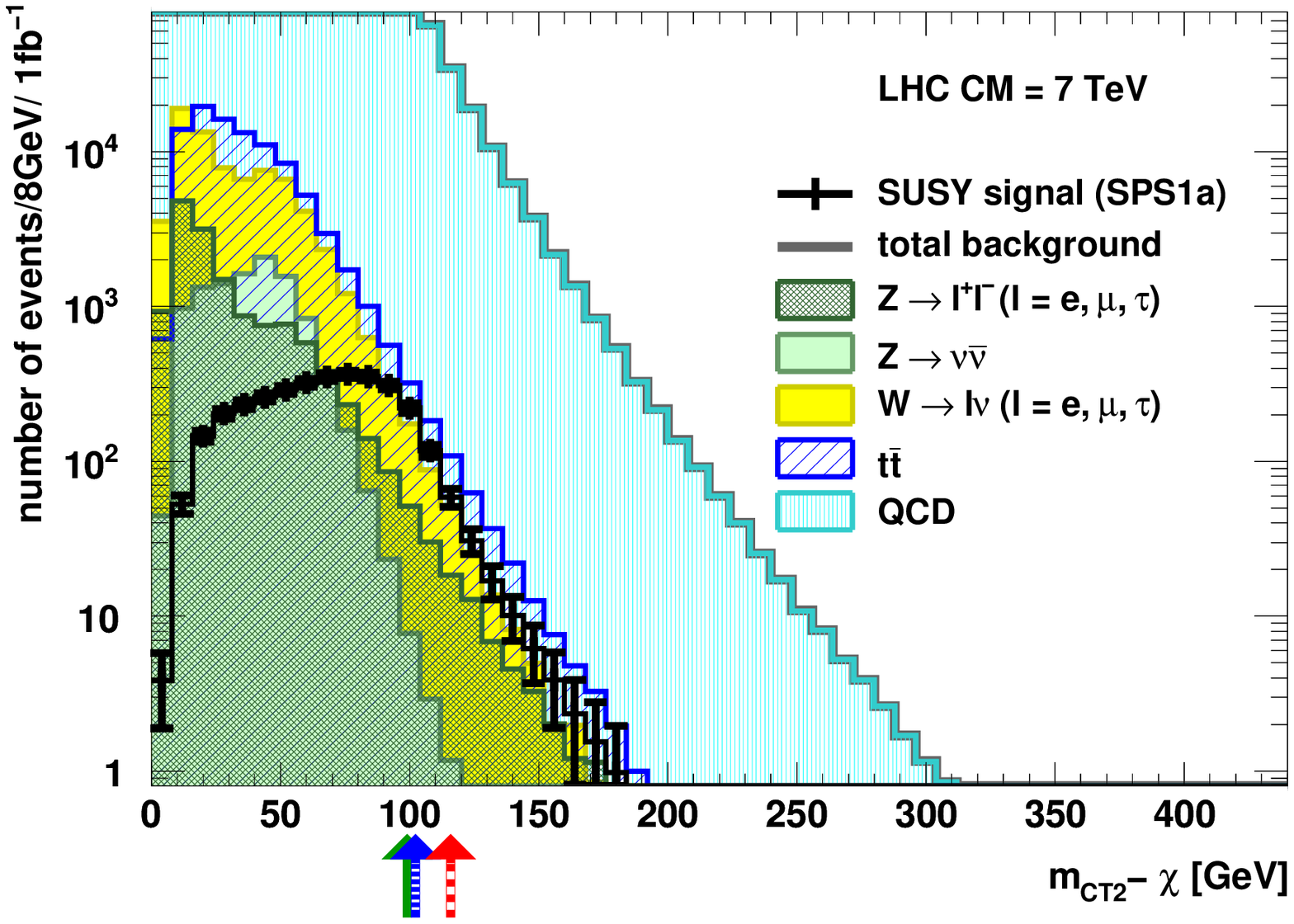}}
  \qquad
  \setcounter{subfigure}{3}
  \subfigure[$m_{T2}(\chi=0)$ distribution with just the dijet cuts applied.]{\label{fig:MT2dijetcuts}\includegraphics[width=77mm,trim=0 21 0 21,clip]{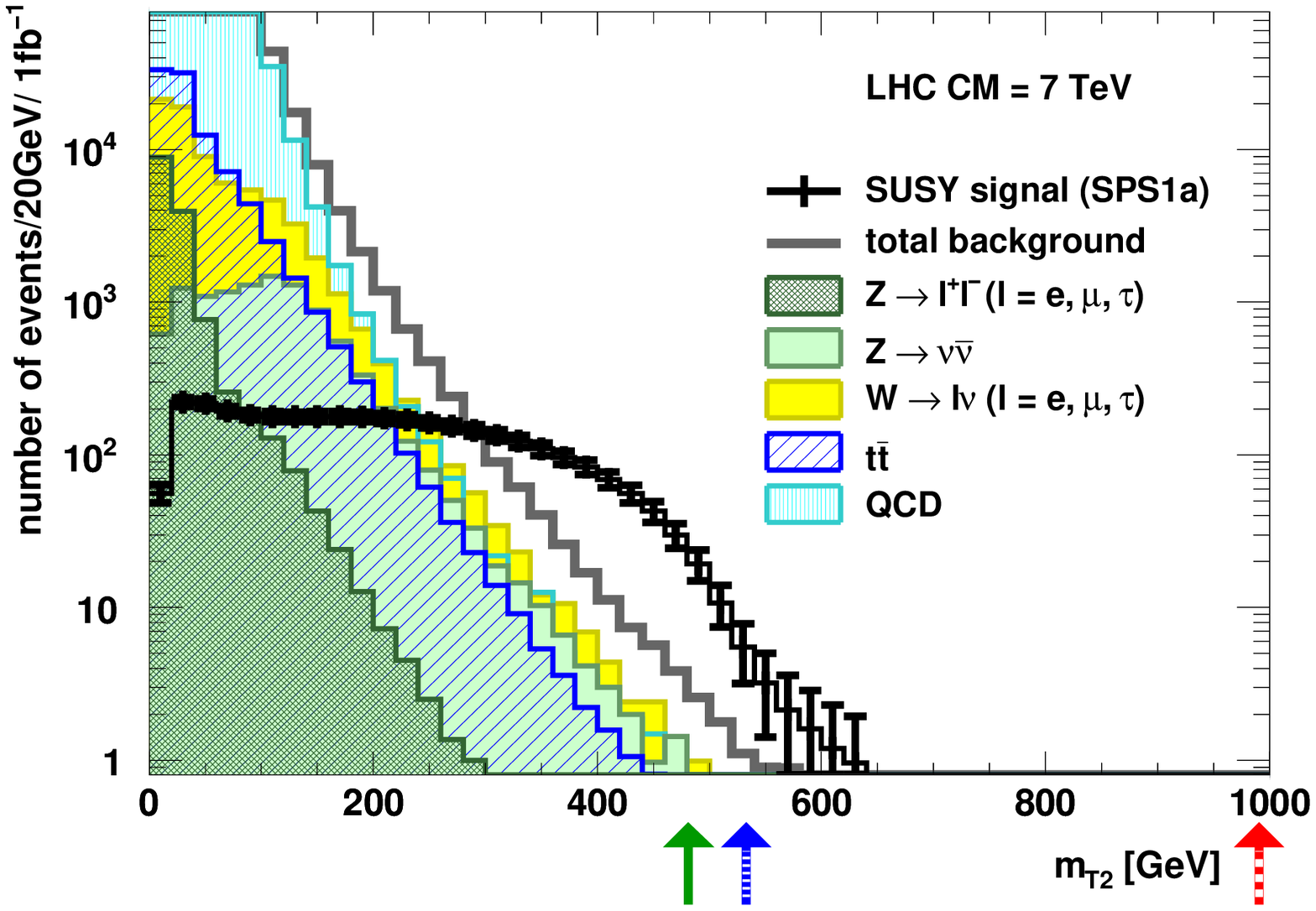}}
  \\ \vspace{-3mm}
  \setcounter{subfigure}{1}
  \subfigure[$m_{CT2}(\chi=300\,\textrm{GeV})$ distribution after the cuts of \cite{ATL-PHYS-PUB-2009-084}.]{\label{fig:MCT2cuts5}\includegraphics[width=77mm,trim=0 21 0 21,clip]{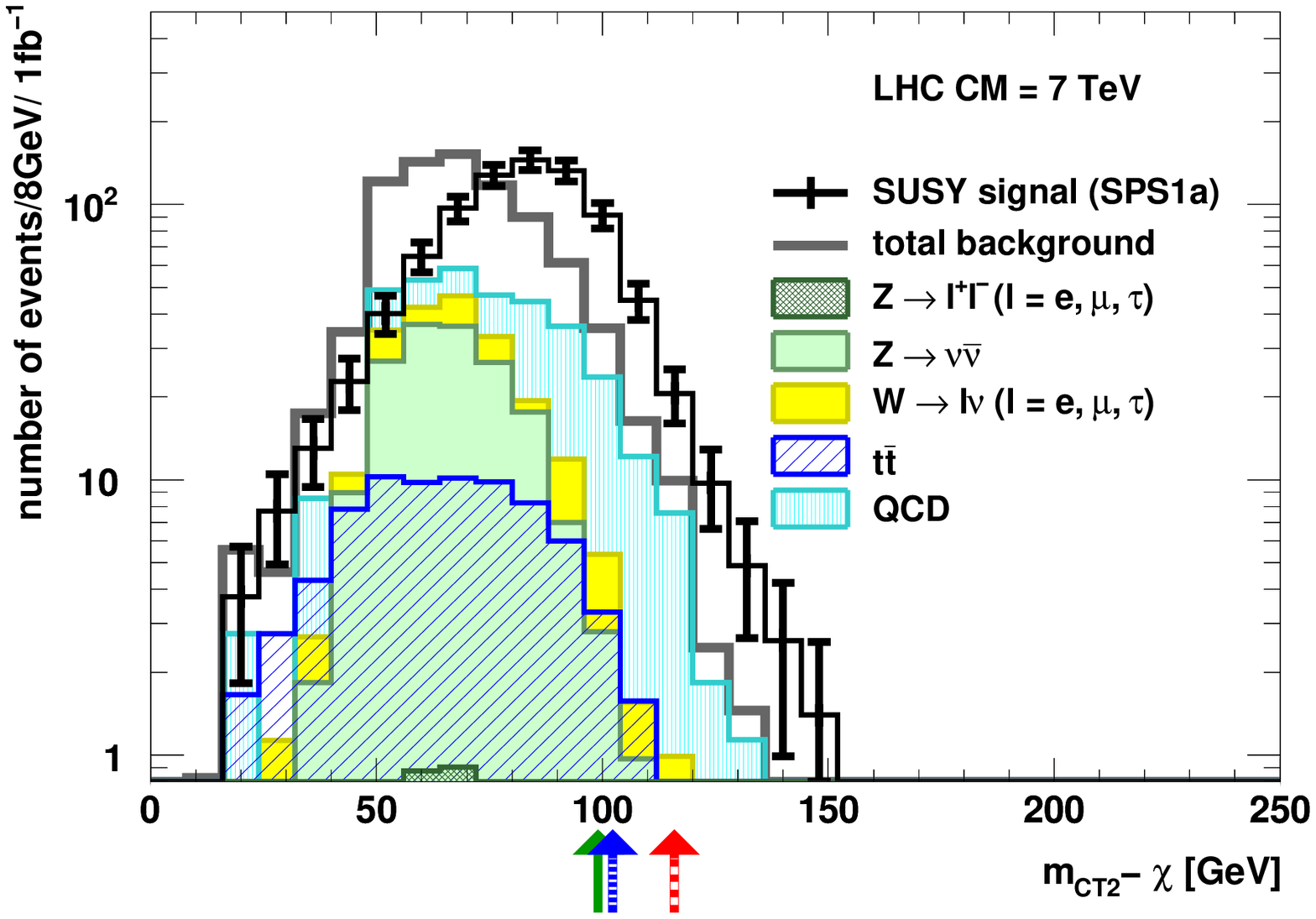}}
  \qquad
  \setcounter{subfigure}{4}
  \subfigure[$m_{\textrm{eff}}$ distribution after the cuts of \cite{ATL-PHYS-PUB-2009-084}.]{\label{fig:MEFFcuts5}\includegraphics[width=77mm,trim=0 21 0 21,clip]{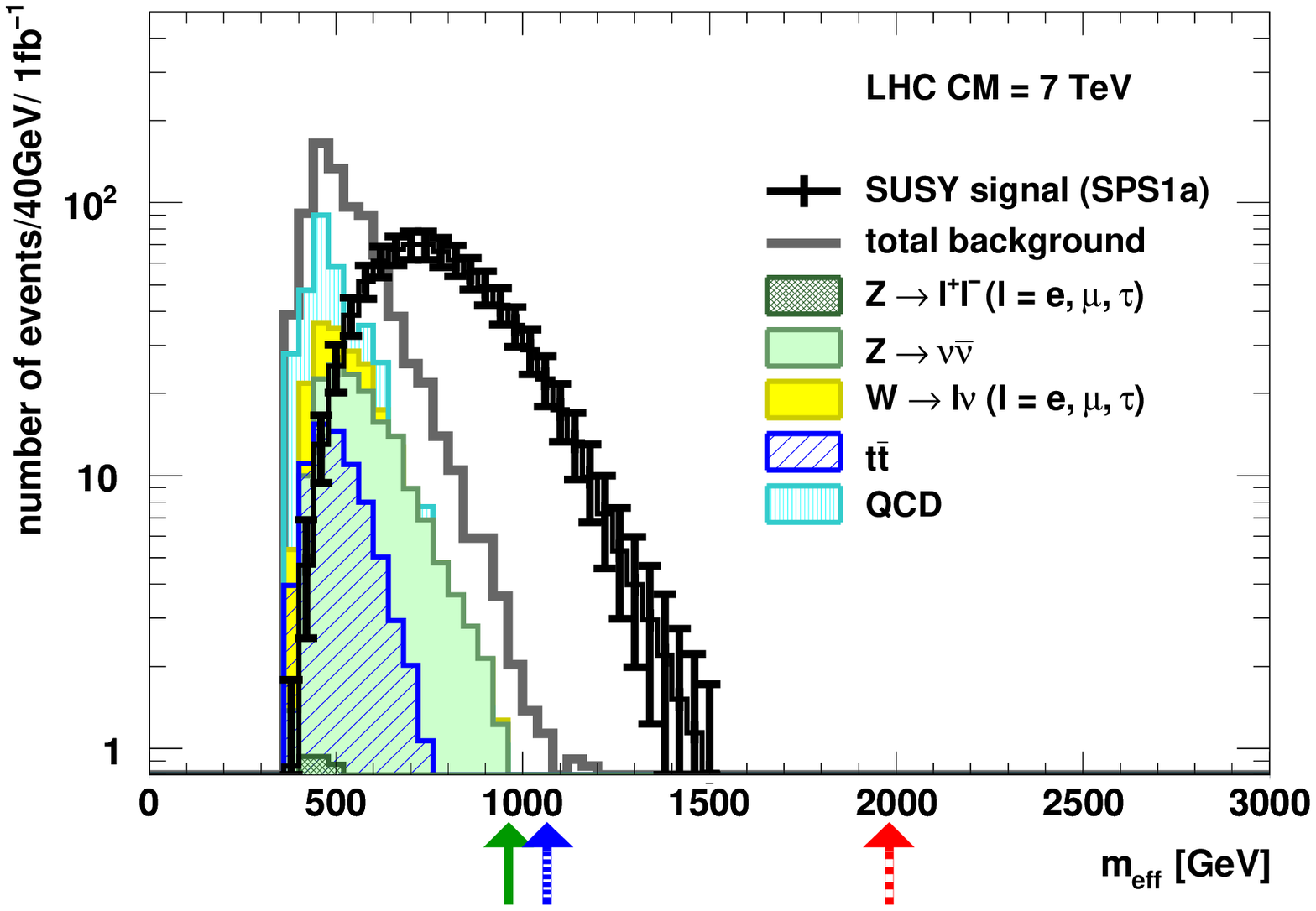}}
  \\ \vspace{-3mm}
  \setcounter{subfigure}{2}
  \subfigure[$m_{CT2}(\chi=300\,\textrm{GeV})$ distribution after the cuts of \cite{ATL-PHYS-PUB-2009-084} and an additional cut requiring $|\delta|<30\,\textrm{GeV}$.]{\label{fig:MCT2cuts5andDelta}\includegraphics[width=77mm,trim=0 21 0 21,clip]{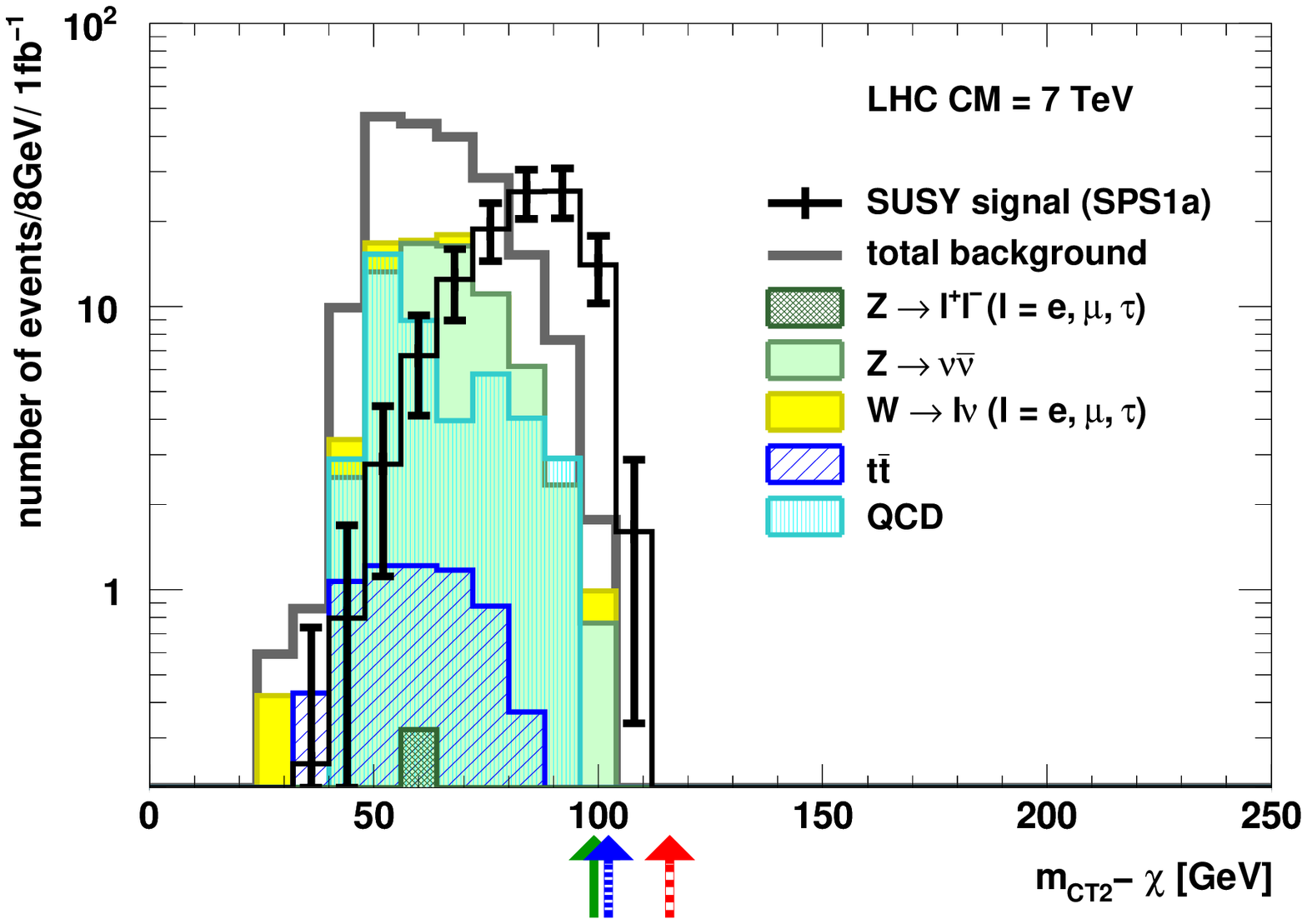}}
  \qquad 
  \setcounter{subfigure}{5}
  \subfigure[$m_{T2}(\chi=0)$ distribution after the cuts of \cite{ATL-PHYS-PUB-2009-084} and an additional cut requiring $|\delta|<30\,\textrm{GeV}$.]{\label{fig:MT2cuts5andDelta}\includegraphics[width=77mm,trim=0 21 0 21,clip]{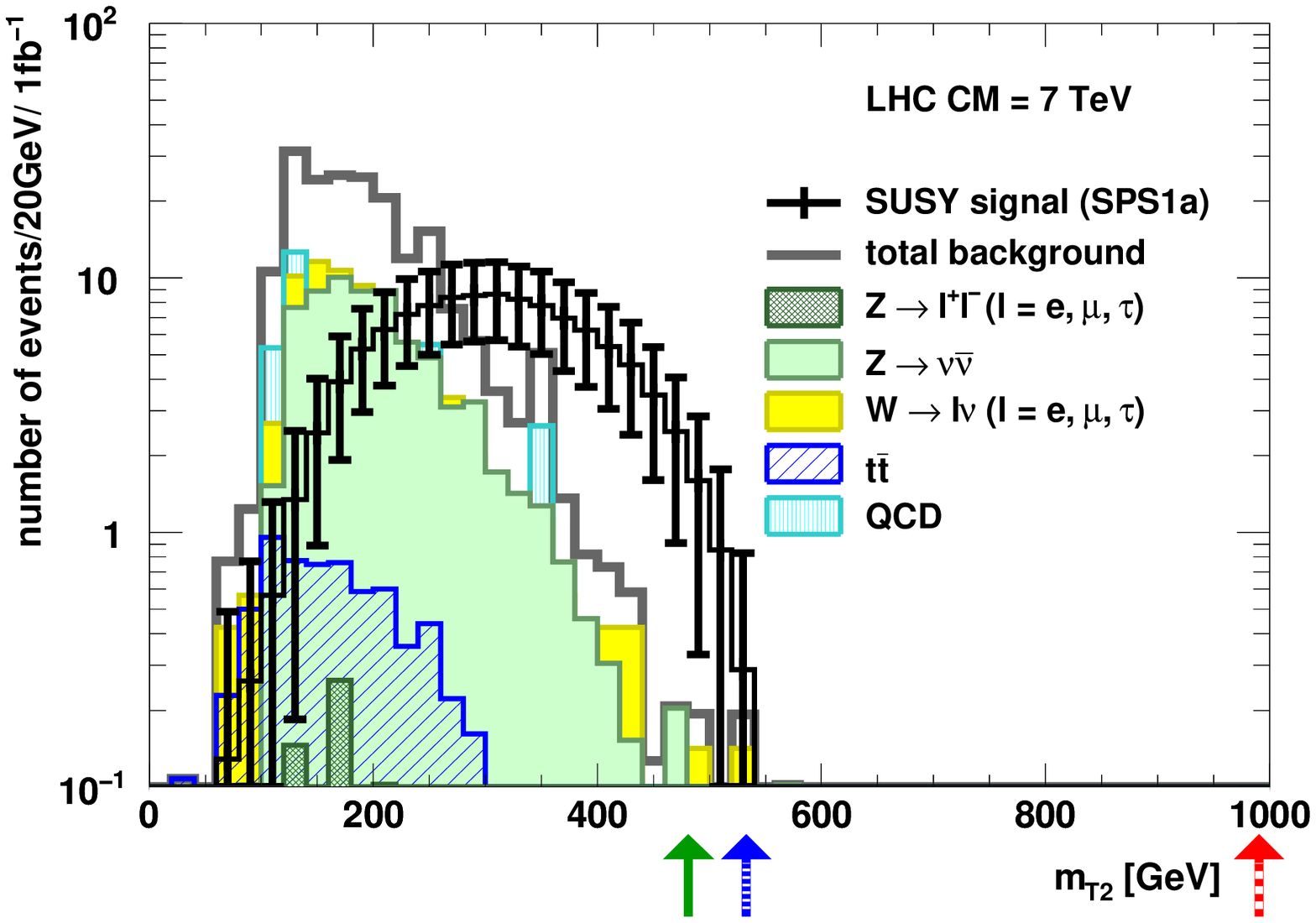}}
  \caption{Results of the simulation described in the text for the SPS1a SUSY benchmark point and various backgrounds. For this signal point the squark masses are in the range $500\lesssim m_{\tilde{q}}\lesssim 600\,$GeV, the gluino mass is close to $600\,$GeV and the lightest neutralino mass is $97\,$GeV. The green (solid), blue (dotted) and red (dashed) arrows along the bottom of the plots show positions of the peaks \eqref{eq:mct2max} or end-points \eqref{eq:mt2max} \eqref{eq:meffpeak} the distributions would be expected to have (under the condition \eqref{eq:noutm}) for processes with masses ($m_Y=500\,$GeV, $m_\chi=97\,$GeV), ($m_Y^{\prime}=1.1m_Y$, $m_\chi$) and ($m_Y^{\prime\prime}=2m_Y$, $m_\chi$) respectively.
(The green and blue arrows lie almost on top of one another in the $m_{CT2}$ plots).
All plots correspond to integrated luminosity of $1\,\textrm{fb}^{-1}$.
}
\end{minipage}
\end{center}
\end{figure*}

As is highlighted in \cite{Cho:2009ve}, a judicious choice of $\chi$ is needed if the distribution of $m_{CT2}$ is to be sensitive to the particle masses.
Cho {\it et al.} show that if one selects a value $\chi\gg|p_0|$ one loses the sharp peak in the $m_{CT2}$ distribution. 

The variation of the endpoint position \eqref{eq:mct2max} with respect to $|p_0|$ is 
\begin{equation}
\frac{\partial m_{CT2}^{\textrm{max}}}{\partial |p_0|}=\frac{{\left(\left(m_{CT2}^{\textrm{max}}\right)^2-\chi^2\right)}^2}{4m_{CT2}^{\textrm{max}}E_T(p_0,\chi)|p_0|^2}.
\end{equation}
so if one chooses the other extreme with $\chi \ll |p_0|$, then
\begin{equation}
\left(m_{CT2}^{\textrm{max}}\right)^2\approx 2\chi^2\left(1-\frac{1}{8}\frac{\chi^2}{|p_0|^2}\right),
\end{equation}
which has very limited sensitivity to the physical parameter $|p_0|$.
Therefore the value of $\chi$ that ought to be chosen should be close to $|p_0|$ (rather than for example the true invisible particle mass). 
For $\chi/|p_0|=\{0.5,\,1,\,1.5\}$, $\frac{\partial m_{CT2}^{\textrm{max}}}{\partial |p_0|}\approx \{0.02,\,0.1,\,0.2\}$ respectively. As was recognised in \cite{Cho:2009ve}, even with a well-tuned $\chi$ the peak position does not vary much with $|p_0|$ so an accurate and precise measurement of  $m_{CT2}^{\textrm{max}}$ is required to constrain $|p_0|$ (and from it the particle masses).

To study the effect that Standard Model backgrounds might have on the method proposed, we use a simulation similar to that described in \cite{Barr:2009wu} including a parameterised detector response typical of a general-purpose LHC detector. {\tt Herwig++~2.4.2} \cite{Bahr:2008tf,Bahr:2008pv} is used to produce samples of the following Standard Model backgrounds; QCD, $t\bar{t}$, $W\rightarrow l\nu$+jets, $Z\rightarrow l^{+}l^{-}$+jets and $Z\rightarrow\nu\nu$+jets for proton-proton collisions at centre of mass energy $7\,$TeV. For a signal we use inclusive production of the SPS1a supersymmetry benchmark point \cite{Allanach:2002nj} with 
the spectrum and decay table calculated by {\tt SPheno~2.2.3} \cite{Porod:2003um}.
Jets are formed and smeared following the same procedure as described in \cite{Barr:2009wu}. 

The expected peaking of the $m_{CT2}(\chi=300\,\textrm{GeV})$ variable can seen in Fig.~\ref{fig:MCT2dijetcuts} after applying only minimal selection cuts requiring at least two jets with $p_T>50\,$GeV and with pseudorapidity, $|\eta|<2.5$. 
We have chosen $\chi=300\,\textrm{GeV}$, close to $p_0$ (which ranges from $240\,\textrm{GeV}\lesssim |p_0|\lesssim 300\,\textrm{GeV}$ depending on which squark mass is used).
In the region of the SUSY signal, it can be seen that the Standard Model backgrounds are large. 

It can also be seen from Fig.~\ref{fig:MCT2dijetcuts} that the dependence of the position of the peak in the $m_{CT2}$ distribution on the physical parameter $|p_0|$ that one is trying to measure is very small. This is shown by the different coloured arrows on that plot indicating expected signal peak positions corresponding to different parent particle masses. Even with large variations in the parent particle mass (up to 100\%) little variation in the peak position is seen. This means that a very precise determination of the peak position would be required in order to get sensitivity to the parent particle mass.\footnote{This is also true for other values of $\chi$.}

In Fig.~\ref{fig:MCT2cuts5} we show the $m_{CT2}(\chi=300\,\textrm{GeV})$ distribution after the application of more selective SUSY cuts  \cite{ATL-PHYS-PUB-2009-084} proposed by ATLAS. The backgrounds appear reduced but are still significant in the signal region, and they have a peaked structure similar to the signal. These backgrounds will therefore have to be well-understood for the successful extraction of the signal shape.


As was pointed out in \cite{Cho:2009ve} it is possible to sharpen the end-point structure of the $m_{CT2}$ distribution by  selecting a subset of events for which the `upstream momentum' is small; i.e. those for which \eqref{eq:noutm} is approximately satisfied.
In Fig.~\ref{fig:MCT2cuts5andDelta} we show the  $m_{CT2}$ distribution after an additional selection requiring 
$|\delta|<30\,\textrm{GeV}$ where the upstream momentum
$\delta = -\slashed{p}_T - p_T^{(1)} - p_T^{(2)}$. 
The $m_{CT2}$ endpoint position becomes better defined when this additional requirement is applied, but at the cost of a factor of $\sim10$ in number of events, and even then the residual backgrounds are still not negligible (given the precision in the endpoint position that is required).

To investigate the effect of the Standard Model background on the precision with which $|p_0|$ can be determined we parameterised separately the signal s and the background b distributions from Fig.~\ref{fig:MCT2cuts5andDelta} by two-part piecewise Gaussian functions, 
\[G(\mu^\textrm{s,b},\sigma_1^\textrm{s,b}): (x>\mu^\textrm{s,b});\  G(\mu^\textrm{s,b},\sigma_2^\textrm{s,b}): (x<\mu^\textrm{s,b}).
\]
 Even with a high cross section SUSY model (such as the SPS1a point shown) and assuming perfect knowledge of the shape of the backgrounds\footnote{Only the signal peak position and the normalisations of the signal and background are allowed to vary in the fit.}, a significant increase in the statistical uncertainty is found when the backgrounds are introduced. We also investigated the case when the SUSY cross-section is a fraction of that shown. For $1\,\textrm{fb}^{-1}$ the statistical precision on $|p_0|$ was \{($\pm 6.8$,$\pm 9.6$), ($\pm 8.0$,$\pm 12$), $(\pm14,\pm22)$\}\,GeV for the cases of: original cross-section without and with backgrounds included, half of the original cross-section without and with backgrounds included, and a quarter of the original cross-section without and with backgrounds included respectively. Uncertainties in the {\em shape} of the background contribution will further increase the detrimental effect of the backgrounds.

These difficulties with the constransverse mass variable are not shared by other variables which have previously been proposed for mass measurement. 
For illustration we compare to distributions of two previously proposed variables. 
The first comparison is against the {\em stransverse mass} $m_{T2}$ \cite{Lester:1999tx,Barr:2003rg},
which is defined by\footnote{
For $m_v=0$ the transverse mass is
\begin{equation}\label{eq:mt2def}
m_T^2(p_T,\slashed{q}_T,\chi)
=\chi^2+2E(p_T,0)E(\slashed{q}_T,\chi)-2 {p}_T\cdot{\slashed{q}}_T.
\end{equation}
Note the negative sign before the inner two-vector product which distinguishes $m_T$ from the {\em con}transverse mass $m_{CT}$ defined in \eqref{eq:mctdef}.}
\begin{eqnarray}
\!\!\!\!\!\!\!\!\!\!m_{T2}(p^{(1)}_T,p^{(2)}_T,\slashed{p}_T,\chi)\equiv \qquad\qquad\qquad\qquad\qquad\nonumber\\
\underset{\sum\slashed{q}^{(i)}_T = \slashed{p}_T}{\textrm{min}}\left(\textrm{max}\left(m_T(p^{(1)}_T,\slashed{q}_T^{(1)}),m_T(p^{(2)}_T,\slashed{q}_T^{(2)})\right)\right).
\end{eqnarray}

In Fig.~\ref{fig:MT2dijetcuts} we show a distribution the $m_{T2}(\chi=0)$ distribution after applying the same basic dijet cuts used for  Fig.~\ref{fig:MCT2dijetcuts}.
The end-point of $m_{T2}(\chi=0\,\textrm{GeV})$ distribution is also dependent on the physical quantity $|p_0|$, and under the condition \eqref{eq:noutm}, 
\begin{equation}m_{T2}^{\textrm{max}}(\chi=0)=2|p_0|\label{eq:mt2max}\end{equation} 
so the comparison is suitable.
From the illustrative arrows it is clear that the end-point of the $m_{T2}$ distribution is much more sensitive to the parent particle mass and that this end-point is in a region where Standard Model backgrounds are suppressed.  

In  Fig.~\ref{fig:MEFFcuts5} we show the distribution of another well-established SUSY mass-scale variable: the {\em effective mass}  \cite{Hinchliffe:1996iu},
\begin{equation}\label{eq:meffdef}
m_{\textrm{eff}} = |{\slashed{p}}_T| + \sum_{i=1,2} |{p}_T^{(i)}| 
\end{equation}
where in this paper the sum is over the two jets with the largest transverse momenta.
Since heavy particles are produced near threshold, the effective mass \cite{Tovey:2000wk} is expected to have an endpoint around
\begin{equation}\label{eq:meffpeak}
m_{\textrm{eff}}^\textrm{max} = 4|p_0|.
\end{equation}
It can be seen from Fig.~\ref{fig:MEFFcuts5} that in contrast to $m_{CT2}$, the $m_{\textrm{eff}}$ endpoint is at a value for which the Standard Model background is small and that its characteristic value changes rapidly as a function of $|p_0|$.

Even after the $\delta$ cut designed to improve the peaked nature of the $m_{CT2}$ endpoint the $m_{T2}$ distribution shows more promise for the signal point investigated (Fig.~\ref{fig:MT2cuts5andDelta}).


In conclusion, while it is true that the sharply peaked nature of the signal in the $m_{CT2}$ distribution may lead to lower systematic errors in the fitting procedure, there are large backgrounds from Standard Model processes which also peak near this region. These backgrounds together with the weak dependence of the peak position on the physical parameter $|p_0|$ suggest that mass measurement using this variable will be difficult. We suggest that for the decay process studied here, other existing mass measurement variables 
show more promising characteristics for SUSY mass measurement.

\bibliographystyle{unsrt}
\bibliography{MT2vsMCT2}
\end{multicols}
\end{document}